\documentclass[a4paper,twoside,11pt]{article}
\usepackage[bindingoffset=0.5cm,textheight=22.5cm,hdivide={2.9cm,*,2.8cm}, vdivide={*,22cm,*}]{geometry}
\usepackage{amsmath}
\usepackage{amsfonts}
\usepackage[dvips]{graphicx}
\usepackage[dvips,bookmarksnumbered=true,breaklinks=true]{hyperref}

\newcommand{\uim}{UV/IR mixing}
\newcommand{\nc}{non-commu\-ta\-tive}
\newcommand{\etal}{{\it et al.}}
\newcommand{\eqnref}[1]{Eqn.~(\ref{#1})}		
\newcommand{\figref}[1]{Fig.~\ref{#1}}			
\newcommand{\secref}[1]{Section~\ref{#1}}		
\newcommand{\appref}[1]{Appendix~\ref{#1}}		
\newcommand{\starco}[2]{\left[#1\stackrel{\star}{,}#2\right]}		
\newcommand{\var}[2]{\frac{\d #1}{\d #2}}				
\newcommand{\pa}{\partial}						
\newcommand{\ri}{{\rm i}}						
\renewcommand{\k}{\tilde{k}}						
\newcommand{\p}{\tilde{p}}						
\newcommand{\q}{\tilde{q}}						
\newcommand{\Dt}{\widetilde{D}}						
\newcommand{\wtsq}{\widetilde{\square}}					
\newcommand{\bc}{\bar{c}}						
\newcommand{\Gam}{\Gamma^{(0)}}						
\newcommand{\Act}{S}
\renewcommand{\a}{\alpha}

\renewcommand{\d}{\delta}
\newcommand{\e}{\epsilon}
\newcommand{\vare}{\varepsilon}

\renewcommand{\th}{\theta}

\renewcommand{\l}{\lambda}
\newcommand{\m}{\mu}
\newcommand{\n}{\nu}

\renewcommand{\r}{\rho}
\newcommand{\s}{\sigma}

\renewcommand{\Xi}{\Xi}

\newcommand{\inv}[1]{\frac{1}{#1}}				

\newcommand{\intx}{\int d^4x}						
\newcommand{\F}{\widetilde{F}}

\newcommand{\wsq}{\widetilde{\square}}
\newcommand{\ig}{{\rm i}g}
\newcommand{\intk}{\int d^4k}
\newcommand{\nok}{k^2+\frac{a'^2}{\tilde{k}^2}} 	
\newcommand{\gnok}[1]{\mathcal{G}(#1)}
\title{\vspace{1.5cm}One-Loop Calculations for a Translation Invariant Non-Commutative Gauge Model}

\author{Daniel N. Blaschke\footnotemark[1]~,
Arnold Rofner\footnotemark[1]~, Manfred Schweda\footnotemark[1]~,\\*[12pt] and Ren\'e I.P. Sedmik\footnotemark[1]}

\date{April 8, 2009}

\graphicspath{{./figures/}}

\begin{document}

\maketitle

\thispagestyle{empty}

\begin{center}
\renewcommand{\thefootnote}{\fnsymbol{footnote}}
\vspace{-0.3cm}\footnotemark[1]Institute for Theoretical Physics,
Vienna University of Technology\\
Wiedner Hauptstrasse 8-10, A-1040 Vienna (Austria)\\[0.3cm]
\ttfamily{E-mail: blaschke@hep.itp.tuwien.ac.at, arofner@hep.itp.tuwien.ac.at, mschweda@tph.tuwien.ac.at, sedmik@hep.itp.tuwien.ac.at}
\vspace{0.5cm}
\end{center}%
\begin{abstract}
In this paper we discuss one-loop results for the translation invariant {\nc} gauge field model we recently introduced in ref.~\cite{Blaschke:2008a}. This model relies on the addition of some carefully chosen extra terms in the action which mix long and short scales in order to circumvent the infamous {\uim}, and were motivated by the renormalizable {\nc} scalar model of Gurau \etal~\cite{Rivasseau:2008a}. 
\end{abstract}

\tableofcontents
\section{Introduction}
\label{sec:intro}
Motivated by the goal of combining quantum field theories and gravity in a consistent manner, numerous models of quantum field theories using non-commuting space-time coordinates have been studied in recent years~\cite{Schroer:2004,Rivasseau:2007a}. Especially, models on flat $\th$-deformed spaces have been considered where the non-commutativity is
implemented by the Weyl-Moyal star product~\cite{Douglas:2001,Szabo:2001},
\begin{align}
[ x^{\mu} \stackrel{\star}{,} x^{\nu} ] \equiv  x^{\mu} \star
x^{\nu} -x^{\nu} \star x^{\mu} = \ri \th^{\mu \nu} \, ,
\end{align}
and where the parameters $\th^{\mu \nu}= - \th^{\nu \mu}$ are real constants. In the following, we assume that the deformation matrix $( \th_{\mu\nu} )$ has the simple block-diagonal form
\begin{align}\label{eq:def-theta}
( \th_{\mu\nu} )
=\th\left(\begin{array}{cccc}
0&1&0&0\\
-1&0&0&0\\
0&0&0&1\\
0&0&-1&0
\end{array}
\right) \, ,  \qquad {\rm with} \ \; \th \in \mathbb{R} \, .
\end{align}
So far, only some {\nc} scalar models on flat Euclidean space have been found to be renormalizable~\cite{Grosse:2004b,Grosse:2008a,Rivasseau:2008a}.
In a recent letter~\cite{Blaschke:2008a}, however, we proposed a promising candidate for a renormalizable {\nc} $U(1)$ gauge field model\footnote{In fact, there are also two other candidates on the market, but these either reduce the degrees of freedom~\cite{Slavnov:2003,Blaschke:2006a} or break translation invariance and have difficulties with non-trivial vacuum configurations~\cite{Wulkenhaar:2007,Grosse:2007,Blaschke:2007b} (although the latter problem might be solved~\cite{Wallet:2008b}).} in Euclidean space where the gauge invariant action was supplemented by an additional term, i.e.
\begin{align}\label{new-action_inv}
\Act_{\text{inv}}&=\intx\left[\inv{4}F^{\mu\nu}\star F_{\mu\nu}
+\inv{4}F^{\mu\nu}\star\frac{a'^2}{D^2\Dt^2}\star F_{\mu\nu}\right]\,,
\end{align}
where
\begin{align}\label{eq:def-fieldtensor}
F_{\mu\nu}&=\partial_\mu A_\nu-\partial_\nu A_\mu -\ig\starco{A_\mu}{A_\nu}\,,\nonumber\\
\inv{\Dt^2}&\equiv(\Dt^2)^{-1\star}\,,\qquad {\rm i.e.} \ \; \inv{\Dt^2}\star\Dt^2=1\,,\nonumber\\
\Dt ^2 &= \Dt^{\mu} \star \Dt_\mu \, ,
\qquad {\rm with} \ \;
\Dt_\mu=\th_{\mu\nu}D^\nu\,,
\end{align}
and $D^\nu$ is the covariant derivative. Formally, the second term in the action is a gauge invariant extension of the one introduced in the renormalizable {\nc} scalar model proposed by Gurau \etal~\cite{Rivasseau:2008a}, which reads
\begin{align}
\Act[\phi]&\equiv \int\limits_{\mathbb{R} ^4} d^4x \left[\inv{2}
\left(\pa^\mu\phi\star\pa_\mu\phi +m^2\phi\star\phi
-\phi\star\frac{a^2}{\wsq}\phi\right)
+\frac{\l}{4!}\,\phi\star\phi\star\phi\star\phi\right]\, .
\end{align}
The $\inv{\wsq}\,$-term\footnote{Notice that from \eqnref{eq:def-theta} follows $\wsq=\th^2\square$.} was included as a non-local counter term for the quadratic IR divergence which arises due to the infamous {\uim} problem of {\nc} quantum field theory. In fact, it is more than a mere counter term: It modifies the propagator in such a way that it tends to zero for small momenta $k^2$, i.e. in momentum space the scalar propagator reads
\begin{align}
G(k)=\inv{k^2+m^2+\frac{a^2}{\k^2}}\,.
\end{align}
It was in fact this new ``damping behaviour'' that motivated the introduction of the gauge field action (\ref{new-action_inv}) above, as we believe it will cure the {\uim} problem\footnote{It is interesting to mention that the same damping behaviour $\lim\limits_{k^2\to0}G_{\mu\nu}(k)=0$ is also obtained for the gluon propagator in regular QCD when one restricts the path integral to the region $\Omega$ inside the first Gribov horizon~\cite{Gribov:1978,Zwanziger:1993}.}. Note, that the term used here differs from the one one would expect as a counter term for the quadratic IR divergence of a {\nc} gauge model~\cite{Putz:2003}. Such a counter term would be of the form
\begin{align}
\int d^4x \, \F\star\inv{(\Dt^2)^2}\star\F \qquad\text{with }\;\F=\th^{\mu\nu}F_{\mu\nu}\,.
\end{align}
However, this counter term does not lead to an ``improved'' propagator in the sense that it damps IR divergences. Therefore the alternative in \eqnref{new-action_inv} was constructed.

%

\section{The Model and its Feynman Rules}
\label{sec:model}
As mentioned in the introduction, the damping property of the $\inv{\wtsq}$ term in the renormalizable {\nc} scalar $\phi^4$ theory introduced by Gurau {\etal}~\cite{Rivasseau:2008a} motivates the search for an analogon in $U(1)$ gauge theory. A corresponding model has been presented in a recent work \cite{Blaschke:2008a}, leading to the gauge fixed action
\begin{align}\label{eq:gauge_action_woB}
\Gam&=\Act_{\text{inv}}
+\Act_{\text{gf}}\,,
\nonumber\\
\Act_{\text{inv}}&=\intx\left[\inv{4}F^{\mu\nu}\star F_{\mu\nu}
+\inv{4}F^{\mu\nu}\star\frac{a'^2}{D^2\Dt^2}\star F_{\mu\nu}\right]
\,,
\nonumber\\
\Act_{\text{gf}}&=
s \intx \, \bc \star \Bigg[
\left( 1+\frac{a'^2}{\square\wsq} \right) \partial^{\mu} A_{\mu}
- \frac{\alpha}{2} b
\Bigg]
\nonumber\\
&=\intx\Bigg[b\star
\left(1+\frac{a'^2}{\square\wsq}\right)\partial^\mu A_\mu-\frac{\alpha}{2}b\star b
-\bc\star\left(1+\frac{a'^2}{\square\wsq}\right)\partial^\mu D_{\mu} c
\Bigg]
\, ,
\end{align}
in Euclidean space, where $\inv{D^2\Dt^2}\star F_{\mu\nu}$ in the gauge invariant part of the action $\Act_{\text{inv}}$ is to be understood as a formal power series in the gauge field $A_\mu$ (for details we refer to Ref.~\cite{Blaschke:2008a}). The field tensor $F_{\mu\nu}$ has already been defined in \eqnref{eq:def-fieldtensor} in the introduction, $b$ is the Lagrange multiplier field implementing the gauge fixing, $c$ and $\bc$ are the ghost and antighost, and $\a$ and $a'$ are dimensionless parameters. As in the corresponding scalar model we expect the new parameter $a'$ to play an essential role in the renormalization procedure which is the reason it has been introduced in the action.

The gauge fixed action (\ref{eq:gauge_action_woB}) is invariant under the BRST transformations
\begin{align}
&sA_\mu=D_\mu c\equiv \partial_\mu
c-\ig\starco{A_\mu}{c}, && s\bc=b ,
\nonumber\\
&sc=\ig{c}\star{c}, && sb=0,
\nonumber\\
& s^2\varphi=0
\quad \mbox{for} \ \; \varphi\in\left\{A_\mu,c,\bc, b \right\}
\, .
\label{eq:BRST_trafos}
\end{align}
As indicated above in (\ref{eq:gauge_action_woB}), the gauge fixing part $\Act_{\text{gf}}$ is BRST exact. Moreover, it follows that $ s\left( \inv{\Dt^2}\star F_{\mu\nu} \right) =\ig\starco{c}{\inv{\Dt^2} F_{\mu\nu}}$, as was previously shown in Ref.~\cite{Blaschke:2008a}.

However, this model incorporates a drawback arising from the fact that the series expansion of the $\inv{D^2\Dt^2}\star F_{\mu\nu}$ term introduces an infinite number of gauge boson vertices. Hence it is not possible to compute a finite sum of contributions in a given loop order. Furthermore, any approximation to the infinite series $\inv{D^2\Dt^2}\star F_{\mu\nu}$ with just a finite number of terms cannot be gauge invariant, and therefore any explicitly calculated (physical) quantities using such an approximation will fail to be gauge invariant as well.

Fortunately, these problems can be circumvented by the introduction of a new antisymmetric field  $B_{\mu\nu}$ of mass dimension two. The respective terms in the action replacing the infinite power series read
\begin{align}
\Act_{\text{inv}}^{(2)}&=\intx\left[a'B_{\mu\nu}\star F_{\mu\nu}-B_{\mu\nu}\star \Dt^2D^2\star B_{\mu\nu}\right]\,,
\label{eq:gauge_action_B_part}
\end{align}
and gauge invariance is given if $B_{\mu\nu}$ transforms covariantly, i.e.
\begin{align}\label{eq:brst-of-B}
sB_{\mu\nu}=\ig\starco{c}{B_{\mu\nu}}.
\end{align}
By reinserting its equation of motion
\begin{align}
\var{\Act_{\text{inv}}^{(2)}}{B_{\rho\s}}&=a'F_{\rho\s}-2\Dt^2 D^2\star B_{\rho\s}=0
\label{eq:eom_B}
\end{align}
into \eqref{eq:gauge_action_B_part} one arrives once more at the original terms:
\begin{align}
\Act_{\text{inv}}^{(2)}&=\intx\left[\left(\frac{a'^2}{2D^2\Dt^2}\star F_{\mu\nu}\right)\star F_{\mu\nu}-\left(\frac{a'^2}{2D^2\Dt^2}\star F_{\mu\nu}\right)\star  \inv{2}F_{\mu\nu}\right]\nonumber\\
&=\intx\left[\inv{4}F_{\mu\nu}\star\frac{a'^2}{D^2\Dt^2}\star F_{\mu\nu} \right]\,.
\end{align}
Note that the latter term vanishes in the limit $a'\to0$ while in the ``improved'' formulation given in \eqnref{eq:gauge_action_B_part} the second term does not. Hence, the newly introduced $B_{\mu\nu}$ is a dynamic field whose existence is independent from the value of the parameter $a'$.

Finally, from \eqnref{eq:gauge_action_woB} with the second term replaced by \eqnref{eq:gauge_action_B_part} the equations of motion for the free theory are\footnote{As usual, the external sources $j^{\phi}=\{j^A_{\mu},j^B_{\mu\nu},j^b,j^c,j^{\bc}\}$ have been introduced via the Legendre transformation $Z^c_{\text{bil}}[j^\phi]=\Act_{\text{bil}}\left[\phi[j^\phi]\right]+j^\phi\phi[j^\phi]$ with $\phi=\{A,B,b,c,\bc\}$.}
\begin{align}
\var{\Act_{\text{bil}}}{A^\nu}
&=-\left(\square\d_{\nu\mu}-\partial_\nu\partial_\mu\right)A^\mu -2a'\pa^\mu B_{\mu\nu} -\left(1+\frac{a'^2}{\square\wsq}\right)\partial_\nu b=-j^A_{\nu} \,,
\nonumber\\
\var{\Act_{\text{bil}}}{B^{\mu\nu}}
&=a'\left(\pa_\mu A_\nu-\pa_\nu A_\mu\right)-2\wsq\square B_{\mu\nu}=-j^B_{\mu\nu}\,,
\nonumber\\
\var{\Act_{\text{bil}}}{b}
&=\left(1+\frac{a'^2}{\square\wsq}\right)\partial^\mu A_\mu -\alpha b =-j^b\,,
\nonumber\\
\var{\Act_{\text{bil}}}{\bc}
&=-\left(\square+\frac{a'^2}{\wsq}\right) c =j^{\bc}\,.
\label{eq:eom_all}
\end{align}
Hence, one finds the propagators
{\allowdisplaybreaks
\begin{subequations}\label{eq:propagators}
\begin{align}
\raisebox{-8pt}{\includegraphics[trim=4.2ex 0 3ex 0,scale=0.8,clip=true]{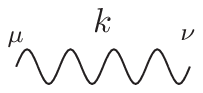}}\hspace{-6pt}&=G^{AA}_{\mu\nu}(k)=\inv{\nok}\left(\d_{\mu\nu}-\frac{k_\mu k_\nu}{k^2}+\alpha\frac{k_\mu k_\nu}{\nok}\right)\,,\label{eq:propagators_AA}\\
\raisebox{-2pt}{\includegraphics[trim=0.2ex 0 0 0,scale=0.8,clip=true]{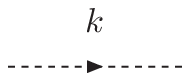}}\hspace{-4pt}&=G^{\bc c}(k)=\frac{-1}{\nok} \,,\\
\raisebox{-8pt}{\includegraphics[trim=4.2ex 0 3ex 0,scale=0.8,clip=true]{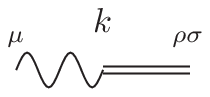}}\hspace{-6pt}&=G^{AB}_{\rho,\s\tau}(k)=\frac{-\ri\, a'}{2}\frac{\left(k_\tau\d_{\rho\s}-k_\s\d_{\rho\tau}\right)}{k^2\k^2\left(\nok\right)},\\
\raisebox{-2pt}{\includegraphics[trim=4.2ex 0 3ex 0,scale=0.8,clip=true]{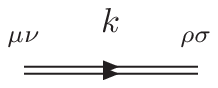}}\hspace{-6pt}&=G^{BB}_{\rho\s,\tau\e}(k)&&\nonumber\\*
 &=\frac{-1}{4k^2\k^2}\!\left[\d_{\rho\tau}\d_{\s\e}\!-\d_{\rho\e}\d_{\s\tau}\!+a'^2\frac{k_\s k_\tau\d_{\rho\e}\!+k_\rho k_\e\d_{\s\tau}\!-k_\s k_\e\d_{\rho\tau}\!-k_\rho k_\tau\d_{\s\e}}{k^2\k^2\left(\nok\right)}\right]\!.
\end{align}
\end{subequations}
The latter two are antisymmetric in the index pairs corresponding to the $B_{\mu\nu}$ fields, i.e.
}
\begin{align}
G^{AB}_{\rho,\s\tau}(k)&=-G^{AB}_{\rho,\tau\s}(k)=-G^{BA}_{\s\tau,\rho}(k)\,,\nonumber\\
G^{BB}_{\rho\s,\tau\e}(k)&=-G^{BB}_{\s\rho,\tau\e}(k)=-G^{BB}_{\rho\s,\e\tau}(k)\,.
\end{align}
Notice furthermore the relations
\begin{align}\label{prop-relations}
2k^2\k^2 G^{AB}_{\rho,\mu\nu}(k)&=\ri a'k_\mu G^{AA}_{\rho\nu}(k)-\ri a'k_\nu G^{AA}_{\rho\mu}(k)\,,\nonumber\\
2k^2\k^2G^{BB}_{\mu\nu,\rho\s}(k)&=\inv{2}\left(\d_{\mu\rho}\d_{\nu\s}-\d_{\mu\s}\d_{\nu\rho}\right)+\ri a'k_\mu G^{BA}_{\rho\s,\nu}(k)-\ri a'k_\nu G^{BA}_{\rho\s,\mu}(k)\,,
\end{align}
which directly follow from the equations of motion for $B_{\m\n}$, \eqnref{eq:eom_B} with the right hand side replaced by the source $-j_{\m\n}^B\neq0$. An interesting question is whether the relations (\ref{prop-relations}) survive on the quantum level, i.e. if they are still valid for the one-loop (or higher-loop) corrected propagators.

The vertices give rather lengthy expressions; thus, they are listed in \appref{app:vertices}. In addition to the usual ones describing interactions between (anti)ghost and gauge bosons, one now also has interactions of gauge bosons with  $B_{\mu\nu}$ fields giving rise to numerous new Feynman diagrams. 

Finally, a few comments on the gauge fixing we have chosen for this model is in order: The purpose of the insertion of the operator $\left(1+\frac{a'^2}{\square\wsq}\right)$ into the gauge fixing part of the action $\Act_{gf}$ in \eqnref{eq:gauge_action_woB} is to further improve the gauge field propagator, i.e. the $\a$-dependent term, as well as the ghost propagator. In the case of $\a=0$ the gauge fixing condition (for vanishing external source $j^b$) is given by
\begin{align}\label{gf-condition}
\left(1+\frac{a'^2}{\square\wsq}\right)\partial^\mu A_\mu=0\,,
\end{align}
which in the limit $a'\to0$ reduces to the regular Lorenz gauge $\partial^\mu A_\mu=0$. However, from \eqnref{eq:propagators_AA} one sees that the gauge boson propagator becomes transversal with respect to $k^\mu$ for $\a=0$ and hence $\partial^\mu A_\mu=0$ is fulfilled even when $a'\neq0$.
%
\section{Power Counting}
\label{sec:counting}
In order to determine the superficial degree of (ultraviolet) divergence of an arbitrary Feynman graph of the present model, we take into account the powers of internal momenta $k$ each Feynman rule contributes and also that each loop integral over 4-dimensional space increases the degree by 4. For example, the gauge boson propagator behaves like $1/k^2$ for large $k$ and therefore reduces the degree of divergence by 2, whereas each ghost vertex (cf. \eqnref{eq:vertices_c} in \appref{app:vertices}) contributes one power of $k$ to the numerator of a graph, hence increasing the degree by one. Continuing these considerations for all other Feynman rules we arrive at
\begin{align}
 d_\gamma&=4L-2I_A-2I_c-5I_{AB}-4I_{BB}+V_c+V_{3A}+3V_{BBA}+2V_{2B2A}+V_{2B3A}\,,
\end{align}
where the $I$ and $V$ denote the number of the various types of internal lines and vertices, respectively (see \eqnref{eq:propagators} and Appendix~\ref{app:vertices}). The number of loop integrals $L$ is given by
\begin{align*}
L&=I_A+I_c+I_{AB}+I_{BB}-\nonumber\\
&\quad-\left(V_c+V_{3A}+V_{4A}+V_{BAA}+V_{BBA}+V_{2B2A}+V_{2B3A}+V_{2B4A}-1\right)\,.
\end{align*}
Furthermore, we take into account the relations
\begin{align}\label{eq:power-counting-relations}
E_{c/\bc}+2I_c&=2V_c\,,\nonumber\\
E_A+2I_A+I_{AB}&=V_c+3V_{3A}+4V_{4A}+2V_{BAA}+V_{BBA}+2V_{2B2A}+3V_{2B3A}+4V_{2B4A}\,,\nonumber\\
E_B+2I_{BB}+I_{AB}&=V_{BAA}+2V_{BBA}+2V_{2B2A}+2V_{2B3A}+2V_{2B4A}\,,\nonumber\\
E_\th&=2I_{AB}+2I_{BB}-2V_{BBA}-2V_{2B2A}-2V_{2B3A}-2V_{2B4A}\,,\nonumber\\
E_{a'}&=I_{AB}+V_{BAA}\,,
\end{align}
between the various Feynman rules describing how they (and how many) can be connected to one another. The $E_{c/\bc}$, $E_A$ and $E_B$ denote the number of external lines of the respective fields whereas $E_\th$ and $E_{a'}$ count the negative powers of $\th$ and positive powers of $a'$ in a graph, respectively. 
Using these relations one can eliminate all internal lines and vertices from the power counting formula. From the last three lines of \eqnref{eq:power-counting-relations} it follows that $E_B+E_\th=E_{a'}$, and  therefore we find two alternative expressions for the power counting, reading
\begin{subequations}\label{eq:counting}
\begin{align}
d_\gamma&=4-E_A-E_{c/\bc}-2E_B-2E_\th\,,\label{eq:counting1}\\
d_\gamma&=4-E_A-E_{c/\bc}-2E_{a'}\,.\label{eq:counting2}
\end{align}
\end{subequations}
In \eqnref{eq:counting1} the superficial degree of divergence is reduced by the number of external legs weighted by the dimension of the respective fields (and parameters). However, this formula can be misleading, since $E_\th$ may also become negative in some graphs. Therefore, in practice, it is more convenient to use \eqnref{eq:counting2}.
%
\section{One-loop Calculations}
\label{sec:one-loop}
%
%
\subsection{Preliminary Considerations}
\label{sec:one-loop_principal}
Consider the simplest integral appearing in non-planar one-loop graphs of the model \eqref{eq:gauge_action_woB}:
\begin{align}
\intk\frac{e^{\ri k\p}}{\nok}\,.
\label{eq:p-2_propag_int}
\end{align}
Superficially, one would expect this integral to diverge quadratically necessitating the introduction of a UV cutoff. However, it is regularized by the phase and no cutoff is needed. This is the usual mechanism leading to {\uim}, and for dimensional reasons the above integral is expected to be $\propto\inv{\p^2}$. However, insertions of such an IR divergent expression into higher loop graphs are regularized by the IR damping behaviour of additional propagators (see also ref.~\cite{Blaschke:2008b}). Hence, the present gauge model is expected to remedy the {\uim} problem.

In order to do the explicit calculation the denominator of \eqref{eq:p-2_propag_int} can be written as
\begin{align*}
\inv{\nok}=\frac{k^2}{\left(k^2+\ri a\right)\left(k^2-\ri a\right)}=\inv{2}\left[\inv{\left(k^2+\ri a\right)}+\inv{\left(k^2-\ri a\right)}\right],
\end{align*}
where $a\equiv a'/\th$. (Note that the rescaled parameter $a$ has mass dimension two.) Furthermore, using Schwinger parametrization one arrives at
\begin{align}
&\sum\limits_{\xi=\pm1}\intk\int\limits_{0}^{\infty}d\alpha \exp\left[-\alpha\left(k^2+\ri\xi a\right)+\ri k\p\right]\nonumber\\
&=\sum\limits_{\xi=\pm1}\int\limits_{0}^{\infty}d\alpha\frac{\pi^2}{\alpha^2}\exp\left[-\alpha\left(\ri\xi a\right)-\frac{\p^2}{4\alpha}\right]\nonumber\\
&=2\pi^2\int\limits_{0}^{\infty}d\alpha\frac{\cos\left(a\alpha\right)}{\alpha^2}e^{-\frac{\p^2}{4\alpha}}
\label{eq:new_param_int}
\end{align}
This integral can be looked up in Ref.~\cite{Gradshteyn:2007} (Eqn.~3.957/2), finally leading to the following combination of modified Bessel functions:
\begin{align}
4\pi^2\sqrt{\frac{a}{\p^2}}\left[e^{\ri\pi/4}K_{-1}\left(e^{\ri\pi/4}\sqrt{a\p^2}\right)+e^{-\ri\pi/4}K_{-1}\left(e^{-\ri\pi/4}\sqrt{a\p^2}\right)\right].
\label{eq:1-loop_simple_result}
\end{align}
For small arguments $z\to0$ the modified Bessel function of first type $K_{-1}=K_{1}$ admits the expansion
\begin{align}
\label{eq:bessel-series}
\inv{z}K_1(z)&=\inv{z^2}+\inv{2}\,\ln z+ \frac{1}{2}
\left(\gamma_E - \ln 2 - \frac{1}{2}\right) + \mathcal{O}(z^2) \,,
\end{align}
where $\gamma_E$ denotes the Euler-Mascheroni constant. Being interested in the IR behaviour of the theory, this expansion can be applied to \eqnref{eq:1-loop_simple_result} for the limit of small external momenta $p$, yielding,
\begin{align}
\frac{8\pi^2}{\p^2}-a\pi^3+\mathcal{O}(\p^2)
\end{align}
The latter result again shows the expected $\inv{\p^2}$ behaviour, which has already been found in scalar theories~\cite{Blaschke:2008b,Rivasseau:2008a}.
%
\subsection{Vanishing Tadpole Graphs}
\label{sec:tadpoles}
\begin{figure}[!ht]
\centering
\includegraphics[scale=0.8]{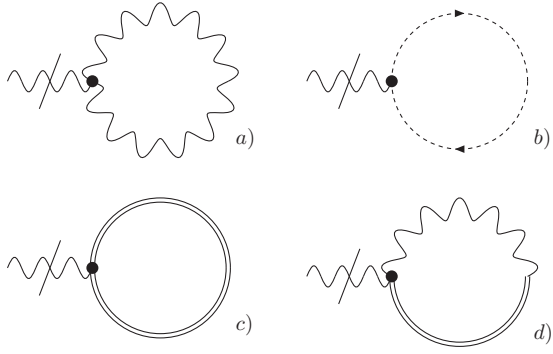}
\caption{One-loop tadpole graphs}
\label{fig:tadpoles}
\end{figure}
\noindent From the Feynman rules \eqnref{eq:propagators} and \eqnref{eq:vertices_1-loop} four possible one-loop tadpole graphs with external gauge boson lines arise. These are depicted in Figure~\ref{fig:tadpoles}.
Each one of the tadpole graphs incorporates a factor 
$\sin\left(\frac{k\p}{2}\right)$ with $p$ and $k$ being the external and internal momenta, respectively. But momentum conservation at the vertices implies $p=0$. Hence, all four graphs vanish.
%
\subsection{Bosonic Vacuum Polarization}
\label{sec:one-loop_vac_pol}
\begin{figure}[!ht]
\centering
\hspace*{-3pt}\includegraphics[trim=10pt 0 10pt 0,scale=0.8,clip=false]{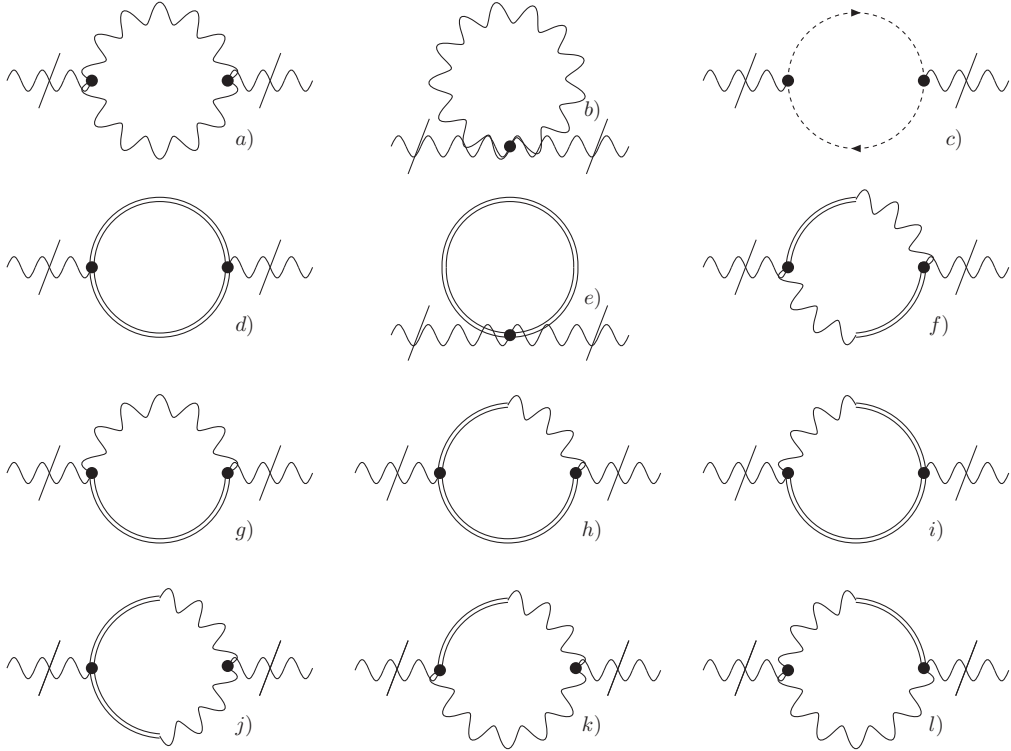}
\caption{Summary of contributions to the one-loop boson vacuum polarization}
\label{fig:1-loop-vac_pol_graphs}
\end{figure}

The present model gives rise to twelve 1PI one-loop graphs with external boson lines. These are collected in \figref{fig:1-loop-vac_pol_graphs} where the first three graphs are already known to appear in theories like QCD.
Computing the respective integrals according to the Feynman rules \eqnref{eq:propagators} and \eqnref{eq:vertices_1-loop} one encounters expressions of the form
\[
\Pi_{\mu\nu}=\intk\, \mathcal{I}_{\mu\nu}(k,p)\sin^2\left(\frac{k\p}{2}\right)\,,
\]
where $\mathcal{I}_{\mu\nu}$ is a function in the external and internal momenta $p$ and $k$, respectively. Details are given in Appendix~\ref{app:one-loop graphs}.

Since we are interested mainly in the IR behaviour of the theory the single-graph results are expanded for small external momenta $p$ according to the rule\footnote{Note, that the phase is not expanded in order to benefit from the regularizing effects in the non-planar parts due to rapid oscillations for large $k$.}
\begin{align}\label{1-l_expansion}
\Pi_{\mu\nu}=&\intk\, \mathcal{I}_{\mu\nu}(p,k) \sin^2\left(\tfrac{k\p}{2}\right)\approx
 \intk\sin^2\left(\tfrac{k\p}{2}\right)\Bigg\{\mathcal{I}_{\mu\nu}(0,k)+p_\rho\left[\partial_{p_\rho}\mathcal{I}_{\mu\nu}(p,k)\right]_{p=0}\nonumber\\
&\hspace{5.7cm}+\frac{p_\rho p_\s}{2}\left[\partial_{p_\rho}\partial_{p_\s}\mathcal{I}_{\mu\nu}(p,k)\right]_{p=0}+\mathcal{O}\left(p^3\right)\Bigg\}.
\end{align}
To lowest order in the expansion (\ref{1-l_expansion}) only five of the graphs depicted in \figref{fig:1-loop-vac_pol_graphs}, namely graphs $a\,$--$\,e$, are found to diverge superficially and read
{\allowdisplaybreaks
\begin{subequations}
\begin{align}
\Pi^{\text{(a)}}_{\mu\nu}&\approx s_a\frac{8g^2}{(2\pi)^4}\intk\frac{\sin^2\left(\frac{k\p}{2}\right)}{\left(\nok\right)^2}\left\{6k_\mu k_\nu+\alpha\, k^2\frac{\left(k^2\d_{\mu\nu}-k_\mu k_\nu\right)}{\nok}\right\},
\\
\Pi^{\text{(b)}}_{\mu\nu}&\approx-s_b \frac{8g^2}{(2\pi)^4}\intk\sin^2\left(\frac{k\p}{2}\right)
\inv{k^2+\frac{a'^2}{\k^2}}
\left[2\d_{\mu\nu}+\frac{k_\mu k_\nu}{k^2}+\a\frac{(k^2\d_{\mu\nu}-k_\mu k_\nu)}{\nok}\right],
\\
\Pi^{\text{(c)}}_{\mu\nu}&\approx-s_c\frac{4g^2}{(2\pi)^4}\intk\sin^2\left(\frac{k\p}{2}\right)\frac{k_\mu k_\nu}{k^4},
\\
\Pi^{\text{(d)}}_{\mu\nu}&\approx s_d\frac{192 g^2}{(2\pi)^4}\intk\sin^2\left(\frac{k\p}{2}\right)\frac{k_\mu k_\nu}{k^4}\left(2+\frac{a'^2\left(\frac{a'^2}{\k^2}-2\right)}{\k^2\left(\nok\right)}\right),
\\
\Pi^{\text{(e)}}_{\mu\nu}&\approx-s_e\frac{24g^2}{(2\pi)^4}\int\mathrm{d}^4k\,\frac{\sin^2(\frac{k\p}{2})}{k^4}[4k_\m k_\n+2k^2\d_{\m\n}]\left(\smash[t]{2-\frac{a'^2}{\k^2\left(\nok\right)}}\right),
\end{align}
\end{subequations}
where the symmetry factors are listed in Table~\ref{tab:symmetry_factors}.
\begin{table}
\centering
\caption{Symmetry factors for the graphs in \figref{fig:1-loop-vac_pol_graphs}}
\begin{tabular}{r@{=}l r@{=}l r@{=}l}
\hline\hline
 \rule[12pt]{1pt}{0pt}$s_a$ & $\inv{2}$ \hspace{1cm} & $s_b$ & $\inv{2}$ \hspace{1cm} & $s_c$ & 1\\[4pt]
 $s_d$ & $\inv{2}$ & $s_e$ & $\inv{2}$ & $s_f$ & 1\\[4pt]
 $s_g$ & 1         & $s_h$ & 1         & $s_i$ & 1\\[4pt]
 $s_j$ & $\inv{2}$ & $s_k$ & 1         & $s_l$ & 1\\[4pt]
\hline\hline
\end{tabular}
\label{tab:symmetry_factors}
\end{table}
The other graphs of \figref{fig:1-loop-vac_pol_graphs} (graphs $f\,$--$\,l$) are found to be finite. This observation is consistent with the power counting formula (\ref{eq:counting}), since graphs $g\,$--$\,l$ come with two overall powers of $a'$, i.e. $E_a=2$, and graph $f$ even has 4 powers of $a'$, i.e. $E_a=4$.

After summing up all these contributions (and neglecting the finite terms) one arrives at the following expression for the quadratic IR divergence:}
\begin{align}\label{transversal-sum}
\Pi^{\text{total}}_{\mu\nu}&=\sum\limits_{j} \Pi^{\text{(j)}}_{\mu\nu}\nonumber\\
&=\frac{4g^2}{(2\pi)^4}\intk\sin^2\left(\frac{k\p}{2}\right)\Bigg[\frac{-1}{\nok}\left(2\d_ {\mu\nu}+\frac{k_\mu k_\nu}{k^2}+\a\frac{(k^2\d_{\mu\nu}-k_\mu k_\nu)}{\nok}\right)-\frac{k_\mu k_\nu}{(k^2)^2}\nonumber\\
&\quad\hspace{4.3cm}+\inv{\left(\nok\right)^2}\left(6k_\mu k_\nu+\a k^2\frac{(k^2\d_{\mu\nu}-k_\mu k_\nu)}{\nok}\right)\nonumber\\
&\quad\hspace{4.3cm}+\frac{12}{k^2}\left(2\frac{k_\mu k_\nu}{k^2}-\d_{\mu\nu}\right)\Bigg]\nonumber\\
&=14\frac{g^2}{\pi^2}\frac{\p_\mu\p_\nu}{(\p^2)^2}+\,\text{finite terms}.
\end{align}
We find a quadratically IR divergent term in the vacuum polarization which is \emph{independent from the gauge parameter $\a$} and \emph{transversal with respect to $p_\mu$}.
One should also note that in the limit $a'\to0$ the $\a$-dependent terms already drop out before integrating over $k$. In fact, $a'$ does not actually play a role in one-loop {\uim} since that effect is dominated by large $k$ for which
\[
 \inv{\nok}\approx\inv{k^2}.
\]
Therefore it is not surprising that the result for the quadratic IR divergence of (\ref{transversal-sum}) does not depend on $a'$. 
Furthermore, it is interesting that the contribution from graphs $d$ and $e$ in \figref{fig:1-loop-vac_pol_graphs} is transversal by itself (as is the contribution from the other three graphs).

Finally, it should be stressed that the zero order terms in \eqnref{transversal-sum} \emph{do not} produce logarithmic IR divergences.
As we will show now, this type of divergence only appears in the second order of the expansion (\ref{1-l_expansion}). (Note, that the first order terms vanish due to the symmetric integration of an odd power in $k$.)
The explicit expressions for the second order terms of the graphs depicted in \figref{fig:1-loop-vac_pol_graphs} with gauge fixing parameter $\a=1$ read:
{\allowdisplaybreaks
\begin{subequations}\label{2nd-order_pre-int}
\begin{align}
\Big(\frac{\partial^2}{\partial p_\r p_\s}\mathcal{I}^{\text{(a)}}_{\mu\nu}\Big)_{p=0}\frac{p_\r p_\s}{2}
& = s_a\frac{4g^2}{(2\pi)^4}\inv{k^4}\Bigg\{10\frac{k_\m k_\n}{k^2}\left(4\frac{(kp)^2}{k^2}-p^2\right)+\d_{\m\n}\left(4\frac{(kp)^2}{k^2}+3p^2\right)\nonumber\\*
&\quad\qquad\quad-2p_\m p_\n-10\frac{(kp)}{k^2}\left(k_\m p_\n+p_\m k_\n\right)+\text{finite terms}\Bigg\},\\
\Big(\frac{\partial^2}{\partial p_\r p_\s}\mathcal{I}^{\text{(b)}}_{\mu\nu}\Big)_{p=0} \frac{p_\r p_\s}{2} & = 0\,,\\
\Big(\frac{\partial^2}{\partial p_\r p_\s}\mathcal{I}^{\text{(c)}}_{\mu\nu}\Big)_{p=0}\frac{p_\r p_\s}{2} & =
s_c\frac{4g^2}{(2\pi)^4}\Bigg\{k_\mu k_\nu\left(\frac{p^2}{k^6}-4\frac{(kp)^2}{k^8}\right)+2p_\mu k_\nu\frac{kp}{k^6}\Bigg\},\\
\Big(\frac{\partial^2}{\partial p_\r p_\s}\mathcal{I}^{\text{(d)}}_{\mu\nu}\Big)_{p=0}\frac{p_\r p_\s}{2} & = s_d\frac{96g^2}{(2\pi)^4}\inv{k^4}\Bigg[p_\m p_\n-4\frac{(kp)}{k^2}\left(k_\m p_\n +p_\m k_\n\right)\nonumber\\*
& \quad\hspace{2.2cm}+4\frac{k_\m k_\n}{k^4}\left(5(kp)^2-k^2p^2\right)+\text{finite terms}\Bigg],\\
\Big(\frac{\partial^2}{\partial p_\r p_\s}\mathcal{I}^{\text{(e)}}_{\mu\nu}\Big)_{p=0}\frac{p_\r p_\s}{2} & = -s_e\frac{24g^2}{(2\pi)^4}\frac{p_\m p_\n}{k^4}\left(2-\frac{a'^2}{\k^2\left(\nok\right)}\right).
\label{eq:B-tad-log}
\end{align}
\end{subequations}
Notice, that the expansion \eqref{1-l_expansion} actually requires $p^2\!\ll k^2$ as becomes clear when looking at the unexpanded expressions containing $\inv{(k+p)^2}$ given in Appendix~\ref{app:one-loop graphs}. Because according to \eqnref{1-l_expansion} we integrate over all $k$ there is a region where this expansion is not valid, but since we are interested in divergences coming from large $k$ this would not be a problem as long as the ``error'' we made was just a finite contribution. Unfortunately, this is not the case in second order of the expansion, i.e. one has logarithmic divergences for large \emph{and} for small $k$. But even if we hadn't made the expansion there would be one logarithmically divergent term present for both large and small $k$: It is the first term in the B-tadpole graph \eqnref{eq:B-tad-log}. However, the sine-squared (cf. \eqnref{1-l_expansion}) cures this divergence, i.e. one has the integral
}
\begin{align}
\intk\frac{\sin^2\left(\frac{k\p}{2}\right)}{k^4}=\inv{2}\intk\frac{\left(1-\cos(k\p)\right)}{k^4}\,.
\end{align}
In order to calculate the planar, i.e. phase independent, part separately one has to introduce a regulator mass $\mu$, and in the end those contributions depending on this cutoff coming from the non-planar sector will exactly cancel the ones from the planar sector. The same technique can be applied to all other terms when considering the expansion above.

Therefore, we conclude:
\begin{itemize}
\item The exact expressions are particularly complicated to calculate explicitly, i.e. some terms require 3 (or possibly more) Schwinger parameters.
\item Even then, we need a regulator mass for one term of the B-tadpole graph.
\item To simplify calculations we hence use the expansion, introduce a regulator mass $\mu$ and expect the limit $\mu\to0$ to be smooth in the sum of planar and non-planar parts.
\end{itemize}
The regulator mass may be introduced in the following way in order to arrive at integrable expressions (namely integrals leading to modified Bessel functions) after the initial Gauss integration:
\begin{align}
\intk \frac{e^{ik\p}}{\left(k^2\right)^2}\;\longrightarrow\;\intk \frac{e^{ik\p}}{\left(k^2+\mu^2\right)^2}\,.
\end{align}
This integral leads to Bessel-integrals of type
\begin{align}
I_\a\equiv\int\limits_0^{\infty}d\a \a^{-1}e^{-\frac{\p^2}{4\a}-\a\mu^2}=2K_0\left(\sqrt{\p^2\mu^2}\right)\,,
\end{align}
which for vanishing regulator reduces to
\begin{align}
\lim\limits_{\mu\to0}I_\a=-2\gamma_E-\ln\left(\frac{\p^2}{4}\right)-\lim\limits_{\mu\to0}\ln(\mu^2)\,.
\end{align}
Returning to \eqnref{2nd-order_pre-int}, we see that we need to solve three types of integrals:
\begin{subequations}
\begin{align}
\intk\frac{e^{\ri k\p}}{\left(k^2+\mu^2\right)^2}&=\pi^2 I_\a\,,\\
\intk\frac{k_\m k_\n e^{\ri k\p}}{\left(k^2+\mu^2\right)^3}&=\frac{\pi^2}{4}\d_{\m\n} I_\a+\text{finite terms}\,,\\
\intk\frac{k_\m k_\n k_\rho k_\s e^{\ri k\p}}{\left(k^2+\mu^2\right)^4}&=\frac{\pi^2}{24}\left(\d_{\m\n}\d_{\rho\s}+\d_{\m\rho}\d_{\n\s}+\d_{\m\s}\d_{\n\rho}\right) I_\a+\text{finite terms}\,,
\end{align}
\end{subequations}
where by ``finite terms'' we mean terms which converge for both $\mu\to0$ and $p\to0$. 
Plugging these formulas into \eqnref{2nd-order_pre-int} we find a result which, as expected, is transversal with respect to $p^\mu$, i.e. $p^\mu \Pi^{\text{log}}_{\m\n}=0$. It is also interesting to note that the second order sum of $B$-dependent graphs (i.e. graphs $d$ and $e$ in \figref{fig:1-loop-vac_pol_graphs}) is also transversal by itself, as indicated below:
\begin{align}\label{log-div_intmed-step}
&\Big(\frac{\partial^2}{\partial p_\r p_\s}\left(\Pi^{\text{(a)}}_{\mu\nu}+\Pi^{\text{(c)}}_{\mu\nu}\right)\Big)_{p=0}\frac{p_\r p_\s}{2}+\Big(\frac{\partial^2}{\partial p_\r p_\s}\left(\Pi^{\text{(d)}}_{\mu\nu}+\Pi^{\text{(e)}}_{\mu\nu}\right)\Big)_{p=0}\frac{p_\r p_\s}{2}=\nonumber\\
&=\frac{2g^2\pi^2}{(2\pi)^4}\Bigg[ \frac{5}{3}\left(p^2\d_{\mu\nu}-p_\m p_\n\right)-2\left(p^2\d_{\mu\nu}-p_\m p_\n\right)\Bigg]\left(I'_\a-I_\a+\text{finite terms}\right)\,,
\end{align}
where the parameter integral $I'_\a$ is given by
\begin{align}
I'_\a\equiv\int\limits_0^{\infty}d\a \a^{-1}e^{-\inv{4\Lambda^2\a}-\a\mu^2}\,,
\end{align}
and $\Lambda$ is an ultraviolet cutoff.
Hence we are led to the final expression where the infrared cutoff $\mu$ drops out in the sum of planar and non-planar parts, as expected:
\begin{align}\label{log-div_final}
\Pi^{\text{log}}_{\m\n}(p)&\equiv\Big(\frac{\partial^2}{\partial p_\r p_\s}\left(\Pi^{\text{(a)}}_{\mu\nu}+\Pi^{\text{(c)}}_{\mu\nu}+\Pi^{\text{(d)}}_{\mu\nu}+\Pi^{\text{(e)}}_{\mu\nu}\right)\Big)_{p=0}\frac{p_\r p_\s}{2}=\nonumber\\
&=\frac{g^2}{24\pi^2}\left(p_\m p_\n-p^2\d_{\m\n}\right)\left(\ln\left(\Lambda^2\p^2\right)+\text{finite terms}\right).
\end{align}
Notice, that we only have a logarithmic divergence in the UV cutoff $\Lambda$ coming from the planar parts and that the result (\ref{log-div_final}) is well-behaved for $p\to0$, i.e. there is no logarithmic infrared divergence in the external momentum.

\section{Conclusion}
In the present paper we have discussed a translation invariant {\nc} gauge model in Euclidean space based on earlier work \cite{Blaschke:2008a,Rivasseau:2008a}. The behaviour regarding divergences of the theory is improved mainly by inserting operators $\left(1+\frac{a'^2}{\square\wtsq}\right)$ in the action, where $a'$ is a free parameter of the theory. As has already been worked out in great detail \cite{Blaschke:2008b} for a similar scalar model these additional factors lead to a damping in all propagators, thereby taming the divergences arising from the UV/IR mixing present in {\nc} models. 
Mathematically, the gauge invariant equivalent of the $\inv{\wsq}$ operator, namely $\inv{D^2\Dt^2}$, acting on $F_{\mu\nu}$ has to be interpreted as an infinite series, leading to an unbounded number of gauge boson vertices. As has been shown in \secref{sec:model} this problem can be circumvented by the introduction of a new antisymmetric field $B_{\mu\nu}$. Initially thought to be a simple Lagrange multiplier, $B_{\mu\nu}$ appears to have its own dynamic properties (c.f. the e.o.m \eqnref{eq:eom_B}), and it remains in the action even for $a'=0$. Nonetheless, it is related to the gauge field by the Ward-like identities \eqref{prop-relations} which suggest a close relation between the new field, and the field strength $F_{\mu\nu}$ of the gauge boson $A_\mu$.

The action \eqref{eq:gauge_action_woB} is invariant with respect to the BRST transformations Eqns.~\eqref{eq:BRST_trafos} and \eqref{eq:brst-of-B}. However, gauge fixing leads to an unconventional condition which reduces to a standard Lorentz gauge for vanishing $\a$. Furthermore, we found a power counting formula \eqref{eq:counting} allowing us to estimate the degree of UV divergence of the model. It appears that powers of the newly introduced parameter $a'$, which are closely tied to the number of external $B_{\mu\nu}$-legs and powers of $\th$, play an essential role.

One-loop calculations show that, due to the phase factors associated with vertices and momentum conservation, tadpole graphs with one external line vanish. Examination of one-loop corrections to the photon propagator show the well-known $\frac{\p_\mu \p_\nu}{(\p^2)^2}$ divergence~\cite{Hayakawa:1999,Ruiz:2000,Blaschke:2005b}. Detailed computation of the divergent part in Eqns.~\eqref{transversal-sum} and \eqref{log-div_final} shows transversality and gauge independence of the quadratic divergence as expected. The latter result is entirely independent from the regulator mass introduced in \secref{sec:one-loop_vac_pol}. 

In a next step we will have to elaborate on the one-loop corrections to the other propagators and the vertices (work in progress) before considering two and higher loop graphs.
It then remains to be shown that the damping mechanism caused by the inserted operators $\inv{D^2\Dt^2}$ and $\inv{\wsq}$ mentioned earlier suffices to render the theory renormalizable at higher loop orders. Furthermore, the physical meaning of the new dynamic field $B_{\mu\nu}$ is not entirely understood up to now, and will be studied in a forthcoming paper.

\section*{Acknowledgments}
The work of D.~N.~Blaschke, A.~Rofner and R.~I.~P.~Sedmik was supported by the ``Fonds zur F\"orderung der Wissenschaftlichen Forschung'' (FWF) under contracts P20507-N1 and P19513-N16.

The authors would like to thank F. Gieres for enlightening discussions.

\appendix
\section{Vertices}
\label{app:vertices}
{\allowdisplaybreaks
\begin{subequations}\label{eq:vertices_1-loop}
\begin{align}
\hspace{-3ex}\raisebox{-20pt}{\includegraphics[scale=0.8,trim=0 5pt 0 5pt,clip=true]{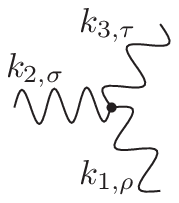}}\hspace{-6pt}&=\widetilde{V}^{3A}_{\rho\s\tau}(k_1, k_2, k_3)\nonumber\\[-10pt]
&=2\ig(2\pi)^4\d^4(k_1+k_2+k_3)\sin\left(\tfrac{k_1\k_2}{2}\right)\times\nonumber\\*
&\quad\times\left[(k_3-k_2)_\rho \d_{\s\tau}+(k_1-k_3)_\s \d_{\rho\tau}+(k_2-k_1)_\tau \d_{\rho\s}\right],\\
\hspace{-3ex}\raisebox{-23pt}{\includegraphics[scale=0.8,trim=0 2pt 0 5pt,clip=true]{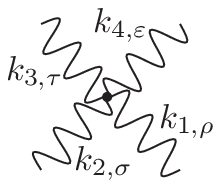}}\hspace{-6pt}&=\widetilde{V}^{4A}_{\rho\s\tau\e}(k_1, k_2, k_3, k_4)\nonumber\\[-10pt]
&=-4g^2(2\pi)^4\d^4(k_1+k_2+k_3+k_4)\times\nonumber\\*
&\quad\times\left[(\d_{\rho\tau}\d_{\s\e}-\d_{\rho\e}\d_{\s\tau})\sin\left(\tfrac{k_1\k_2}{2}\right)\sin\left(\tfrac{k_3\k_4}{2}\right)\right.\nonumber\\*
&\quad\left.+(\d_{\rho\s}\d_{\tau\e}-\d_{\rho\e}\d_{\s\tau})\sin\left(\tfrac{k_1\k_3}{2}\right)\sin\left(\tfrac{k_2\k_4}{2}\right)\right.\nonumber\\*
&\quad\left.+(\d_{\rho\s}\d_{\tau\e}-\d_{\rho\tau}\d_{\s\e})\sin\left(\tfrac{k_2\k_3}{2}\right)\sin\left(\tfrac{k_1\k_4}{2}\right)\right],\\
\hspace{-3ex}\raisebox{-23pt}{\includegraphics[scale=0.8,trim=0 2pt 0 5pt,clip=true]{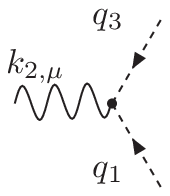}}&=\widetilde{V}^{c}_\mu(q_1, k_2, q_3)\nonumber\\[-15pt]
&=-2\ig(2\pi)^4\d^4(q_1+k_2+q_3)q_{3\mu}\left(1+\frac{a'^2}{(q_3)^2(\tilde{q}_3)^2}\right)\sin\left(\tfrac{q_1\q_3}{2}\right),\label{eq:vertices_c}\\
\hspace{-3ex}\raisebox{-23pt}{\includegraphics[scale=0.8,trim=0 2pt 0 5pt,clip=true]{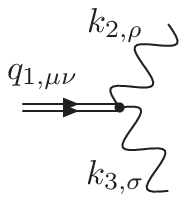}}\hspace{-6pt}&=\widetilde{V}^{BAA}_{\mu\nu,\rho\s}(q_1, k_2, k_3)\nonumber\\[-15pt]
&=2ga'(2\pi)^4\d^4(q_1+k_2+k_3)\left(\d_{\mu\rho}\d_{\nu\s}-\d_{\mu\s}\d_{\nu\rho}\right)\sin\left(\tfrac{k_2\k_3}{2}\right),\\
\hspace{-3ex}\raisebox{-23pt}{\includegraphics[scale=0.8,trim=0 2pt 0 5pt,clip=true]{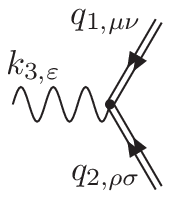}}\hspace{-2pt}&=\widetilde{V}^{BBA}_{\mu\nu,\rho\s,\e}(q_1, q_2, k_3)\nonumber\\[-10pt]
&=-2\ig\th^2(2\pi)^4\d^4(q_1+q_2+k_3)\left(\d_{\mu\rho}\d_{\nu\s}-\d_{\mu\s}\d_{\nu\rho}\right)\times\nonumber\\*
&\quad\,\times\left((q_1)^2+(q_2)^2\right)\left(q_1-q_2\right)_\e\sin\left(\tfrac{q_1\q_2}{2}\right)\,,
\end{align}
\vspace*{-35pt}
\begin{align}
\hspace{4ex}&\raisebox{-23pt}{\includegraphics[scale=0.8,trim=0 5pt 0 5pt,clip=true]{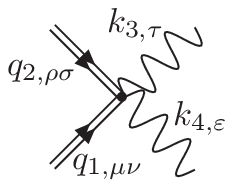}}\hspace{-14pt}=\widetilde{V}^{2B2A}_{\mu\nu,\rho\s,\tau\e}(q_1, q_2, k_3,k_4)\nonumber\\*
&=4g^2\th^2(2\pi)^4\d^4(q_1+q_2+k_3+k_4)\left(\d_{\mu\rho}\d_{\nu\s}-\d_{\mu\s}\d_{\nu\rho}\right)\times\nonumber\\*
&\;\times\!\bigg\{\!\left[k_{3,\tau}k_{4,\vare}\!+2\left(q_{1,\tau}k_{4,\vare}\!+q_{2,\vare}k_{3,\tau}\right)\!+4q_{1,\tau}q_{2,\vare}\!-\d_{\vare\tau}\!\left({q_1}^2+{q_2}^2\right)\right]\sin\!\big(\tfrac{q_1\k_3}{2}\!\big)\sin\!\big(\tfrac{q_2\k_4}{2}\!\big)\nonumber\\*
&\quad+\!\left[k_{3,\tau}k_{4,\vare}\!+2\left(q_{2,\tau}k_{4,\vare}\!+q_{1,\vare}k_{3,\tau}\right)\!+4q_{1,\vare}q_{2,\tau}\!-\d_{\vare\tau}\!\left({q_1}^2+{q_2}^2\right)\right]\sin\!\big(\tfrac{q_1\k_4}{2}\!\big)\sin\!\big(\tfrac{q_2\k_3}{2}\!\big)\bigg\},
\end{align}
\begin{align}
&\raisebox{-23pt}{\includegraphics[scale=0.8]{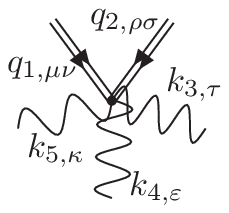}}\hspace{-6pt}=\widetilde{V}^{2B3A}_{\mu\nu,\rho\s,\tau\e\kappa}\,(q_1, q_2, k_3,k_4,k_5)\nonumber\\*
&=-8\ig^3\th^2(2\pi)^4\d^4(q_1+q_2+k_3+k_4+k_5)\left(\d_{\mu\rho}\d_{\nu\s}-\d_{\mu\s}\d_{\nu\rho}\right)\times\nonumber\\*
&\quad\times\Bigg\{\left[k_{3}+2q_{1}\right]_\tau\d_{\e\kappa}\sin\!\big(\tfrac{k_3\q_1}{2}\!\big)\Bigg[\sin\!\big(\tfrac{k_5\q_2}{2}\!\big)\sin\!\big(\tfrac{k_4(\k_5+\q_2)}{2}\!\big)+(k_4\leftrightarrow k_5)\Bigg]\nonumber\\*
&\quad\quad+\left[k_4+2q_{1}\right]_\e\d_{\tau\kappa}\sin\!\big(\tfrac{k_4\q_1}{2}\!\big)\Bigg[\sin\!\big(\tfrac{k_5\q_2}{2}\!\big)\sin\!\big(\tfrac{k_3(\k_5+\q_2)}{2}\!\big)+(k_5\leftrightarrow k_3)\Bigg]\nonumber\\*
&\quad\quad+\left[k_5+2q_{1}\right]_\kappa\d_{\tau\e}\sin\!\big(\tfrac{k_5\q_1}{2}\!\big)\Bigg[\sin\!\big(\tfrac{k_3\q_2}{2}\!\big)\sin\!\big(\tfrac{k_4(\k_3+\q_2)}{2}\!\big)+(k_3\leftrightarrow k_4)\Bigg]\nonumber\\*
&\quad\quad+(q_1\leftrightarrow q_2)\Bigg\},\\
&\raisebox{-23pt}{\includegraphics[scale=0.8]{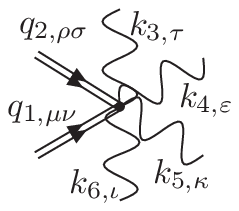}}\hspace{-6pt}=\widetilde{V}^{2B4A}_{\mu\nu,\rho\s,\tau\e\kappa\iota}\,(q_1, q_2, k_3,k_4,k_5,k_6)\nonumber\\*
&=4g^4\th^2(2\pi)^4\d^4(q_1+q_2+k_3+k_4+k_5+k_6)\left(\d_{\mu\rho}\d_{\nu\s}-\d_{\mu\s}\d_{\nu\rho}\right)\times\nonumber\\*
&\quad\times\!\Bigg\{2\d_{\tau\e}\d_{\kappa\iota}\Big[\sin\!\big(\tfrac{k_4\q_1}{2}\big)\sin\!\big(\tfrac{k_3(\k_4+\q_1)}{2}\big)\sin\!\big(\tfrac{k_6\q_2}{2}\big)\sin\!\big(\tfrac{k_5(\k_6+\q_2)}{2}\big)+(k_3\leftrightarrow k_4)+(k_5\leftrightarrow k_6)\Big]\nonumber\\*
&\quad\quad+\d_{\tau\kappa}\d_{\e\iota}\Big[\sin\!\big(\tfrac{k_5\q_1}{2}\!\big)\sin\!\big(\tfrac{k_3(\k_5+\q_1)}{2}\!\big)\sin\!\big(\tfrac{k_6\q_2}{2}\!\big)\sin\!\big(\tfrac{k_4(\k_6+\q_2)}{2}\!\big)+(k_3\leftrightarrow k_5)+(k_4\leftrightarrow k_6)\Big]\nonumber\\[7pt]
&\quad\quad+\d_{\tau\iota}\d_{\kappa\e}\Big[\sin\!\big(\tfrac{k_6\q_1}{2}\!\big)\sin\!\big(\tfrac{k_3(\k_6+\q_1)}{2}\!\big)\sin\!\big(\tfrac{k_4\q_2}{2}\!\big)\sin\!\big(\tfrac{k_5(\k_4+\q_2)}{2}\!\big)+(k_3\leftrightarrow k_6)+(k_5\leftrightarrow k_4)\Big]\nonumber\\*
&\quad\quad+(q_1\leftrightarrow q_2)\Bigg\}.
\end{align}
\end{subequations}
}

\section{One-loop graphs}
\label{app:one-loop graphs}
Using the abbreviation 
\begin{align}
\gnok{k}\equiv \left(k^2+\frac{a'^2}{\k^2}\right)\,,
\end{align}
the full expressions for the graphs depicted in \figref{fig:1-loop-vac_pol_graphs} including the symmetry factors listed in Table~\ref{tab:symmetry_factors} are given by:
{\allowdisplaybreaks
\begin{subequations}
\begin{align}
\Pi^{\text{(a)}}_{\mu\nu}&=s_a\frac{4g^2}{(2\pi)^4}\intk\frac{\sin^2\left(\frac{k\p}{2}\right)}{\gnok{k}\gnok{k+p}}
\Bigg\{\frac{k_\mu k_\nu}{k^2}\Bigg(11k^2-p^2+2kp+\frac{(p^2-k^2)^2}{(k+p)^2}\Bigg)\nonumber\\*
&\quad+\d_{\mu\nu}\Bigg[\a\Bigg(\frac{(k^2+2kp)^2}{\gnok{k}}+\frac{(k^2-p^2)^2}{\gnok{k+p}}\Bigg)\!+k^2+5p^2-2kp-\frac{(k^2-p^2)^2}{(k+p)^2}-4\frac{(kp)^2}{k^2}\Bigg]\nonumber\\*
&\quad+\a k_\mu k_\nu \Bigg(\frac{p^2-k^2-2kp}{\gnok{k}}-
\frac{\frac{p^4}{(k+p)^2}\left(\frac{a'^2}{(\k+\p)^2}-(k+p)^2(\a -1)\right)}{\gnok{k}\gnok{k+p}}-\frac{(k^2-p^2)^2}{k^2\gnok{k+p}}\Bigg)\nonumber\\*
&\quad +p_\mu p_\nu \frac{\left(\frac{a'^2}{(\k+\p)^2}-(k+p)^2(\a -1)\right)}{\left(\gnok{k+p}\right)(k+p)^2}\left(\frac{k^2p^2+(kp)^2-2k^4}{k^2}-\a\frac{(kp)^2}{\gnok{k}}\right)\nonumber\\*
&\quad +p_\mu p_\nu\left(-3+\a\frac{k^2}{\gnok{k}}\right)+(k_\mu p_\nu+ p_\mu k_\nu)\left(3\frac{kp+2k^2}{k^2}-\a\frac{3kp+k^2}{\gnok{k}}\right)\nonumber\\*
&\quad+(k_\mu p_\nu+ p_\mu k_\nu)(kp)\frac{\left(\frac{a'^2}{(\k+\p)^2}-(k+p)^2(\a -1)\right)}{\gnok{k+p}(k+p)^2}\left(\frac{k^2-p^2}{k^2}+\a\frac{p^2}{\gnok{k}}\right)\Bigg\},
\\
\Pi^{\text{(b)}}_{\mu\nu}&=-s_b \frac{8g^2}{(2\pi)^4}\intk\sin^2\left(\frac{k\p}{2}\right)
\inv{\gnok{k}}
\left[2\d_{\mu\nu}+\frac{k_\mu k_\nu}{k^2}+\a\frac{(k^2\d_{\mu\nu}-k_\mu k_\nu)}{\gnok{k}}\right],
\\
\Pi_{\mu\nu}^{\text{(c)}}&=-s_c\frac{4g^2}{(2\pi)^4}\intk\frac{k_\mu(k+p)_\nu}{k^2(k+p)^2}\sin^2\left(\frac{k\p}{2}\right),
\\
\Pi^{\text{(d)}}_{\mu\nu}&= s_d\frac{4g^2\th^4}{(2\pi)^4}\intk\sin^2\left(\frac{k\p}{2}\right)\frac{\left[(k+p)^2+k^2\right]^2(2k+p)_\mu(2k+p)_\nu}{k^2\k^2(k+p)^2(\k+\p)^2}\times\nonumber\\*
&\quad\quad\times\Bigg[6-\frac{3a'^2}{\k^2\gnok{k}}-\frac{3a'^2}{(\k+\p)^2\gnok{k+p}}+\frac{a'^4\left(k^2(k+p)^2+2[k(k+p)]^2\right)}{k^2\k^2(k+p)^2(\k+\p)^2\gnok{k}\gnok{k+p}}\Bigg],
\\
\Pi^{\text{(e)}}_{\mu\nu}&=-s_e\frac{24g^2}{(2\pi)^4}\int\mathrm{d}^4k\,\frac{\sin^2(\frac{k\p}{2})}{k^4}[p_\m p_\n+4k_\m k_\n+2k^2\d_{\m\n}]\left[\smash[t]{2-\frac{a'^2}{\k^2\gnok{k}}}\right],
\\
\Pi^{\text{(f)}}_{\mu\nu}&=s_f\frac{4\,a'^4g^2}{(2\pi)^4}\intk\sin^2\left(\frac{k\p}{2}\right)\frac{3k_\mu k_\nu+2k_\mu p_\nu+k_\nu p_\mu}{k^2\k^2(k+p)^2(\k+\p)^2\gnok{k}\gnok{k+p}},
\\
\Pi^{\text{(g)}}_{\mu\nu}&=s_g\frac{4\,a'^2g^2}{(2\pi)^4}\intk\frac{\sin^2\left(\frac{k\p}{2}\right)}{k^2\k^2\gnok{k+p}}\Bigg\{2\d_{\mu\nu}+\frac{(k+p)_\mu(k+p)_\nu}{(k+p)^2}\nonumber\\*
&\quad+\frac{a'^2}{k^2\k^2\gnok{k}}\Big[\d_{\mu\nu}\left(\tfrac{[k(k+p)]^2}{(k+p)^2}-k^2\right)-k_\mu k_\nu -\tfrac{k(k+p)}{(k+p)^2}\left(2k_\mu k_\nu+k_\mu p_\nu+p_\mu k_\nu\right)\Big]\nonumber\\*
&\quad+\frac{\a}{\gnok{k+p}\gnok{k}}\Bigg(\d_{\mu\nu}\left[k^2(k+p)^2-a'^2\frac{(kp)^2-k^2p^2}{k^2\k^2}\right]-k^2(k_\mu + p_\mu)(k_\nu + p_\nu)\nonumber\\*
&\quad\hspace{3.2cm} -\frac{a'^2}{k^2\k^2}\left(k^2 p_\mu p_\nu +p^2 k_\mu k_\nu -(kp)(k_\mu p_\nu+ p_\mu k_\nu)\right)\Bigg)\Bigg\},
\\
\Pi^{\text{(h+i)}}_{\mu\nu}&=s_{\text{h}}\frac{4\,a'^2g^2}{(2\pi)^4}\intk\frac{\sin^2\left(\frac{k\p}{2}\right)}{\k^2(k+p)^2\gnok{k}}\left(\inv{k^2}+\inv{(k+p)^2}\right)
 \left(2k_\mu + p_\mu\right)\times\nonumber\\*
 &\quad\quad\times\left[3k_\n - a'^2\frac{k_\nu[(k+p)^2+2k(k+p)]+2 p_\nu[k(k+p)]}
 {(k+p)^2(\k+\p)^2\gnok{k+p}}\right] +\mu\leftrightarrow\nu ,
\\
\Pi^{\text{(j)}}_{\mu\nu}&=s_j\frac{4a'^2g^2}{(2\pi)^4}\intk\sin^2\left(\frac{k\p}{2}\right)
\left(\inv{(k+p)^2}+\inv{k^2}\right)\times\nonumber\\*
&\qquad\times\frac{(2k+p)_\mu\left[(6k^2+6kp+2p^2)k_\nu+(3k^2+kp)p_\nu
\right]}{\k^2(k+p)^2\gnok{k}\gnok{k+p}},
\\
\Pi^{\text{(k+l)}}_{\mu\nu}&=s_{\text{k}} \frac{4\,a'^2g^2}{(2\pi)^4}\intk\frac{\sin^2\left(\frac{k\p}{2}\right)}{k^2\k^2\gnok{k}\gnok{k+p}}\Bigg
\{3k_\mu k_\nu+2 p_\mu k_\nu + k_\mu p_\nu \nonumber\\*
&\quad+\d_{\mu\nu}\left[k(k-p)+k(k+p)\frac{(p^2-k^2)}{(k+p)^2}-\a\frac{k(k+p)(p^2-k^2)}{\gnok{k+p}}\right]\nonumber\\*
&\quad+\frac{1}{(k+p)^2}\Big(k(k+p)(k_\mu k_\nu-p_\mu p_\nu)+(p^2+2k^2+3(kp))(k_\mu k_\nu + k_\mu p_\nu)\Big)\nonumber\\*
&\quad -\frac{\a}{\gnok{k+p}}
\Big(k(k+p)(k_\mu k_\nu-p_\mu p_\nu)+(p^2+2k^2+3(kp))(k_\mu k_\nu + k_\mu p_\nu)\nonumber\\*
&\quad\hspace{2.3cm}-(k+p)^2(2k_\mu k_\nu + k_\mu p_\nu) \Big)
+\mu\leftrightarrow\nu \Bigg\}\,.
\end{align}
\end{subequations}


\end{document}